# Choosing Together: How Dyadic Negotiation Shapes Adaptive Kitchen Design Preferences for Older Adults with Cognitive Impairment and Their Care Partners

IBRAHIM BILAU

School of Building Construction, Georgia Institute of Technology, Atlanta, Georgia, United States

SimTigrate Design Center, Georgia Institute of Technology, Atlanta, Georgia, United States

ABDURRAHMAN BARU

School of Building Construction, Georgia Institute of Technology, Atlanta, Georgia, United States

SimTigrate Design Center, Georgia Institute of Technology, Atlanta, Georgia, United States

STACIE SMITH

School of Architecture, Georgia Institute of Technology, Atlanta, Georgia, United States

HUI CAI

School of Architecture, Georgia Institute of Technology, Atlanta, Georgia, United States

SimTigrate Design Center, Georgia Institute of Technology, Atlanta, Georgia, United States

EUNHWA YANG

School of Building Construction, Georgia Institute of Technology, Atlanta, Georgia, United States

SimTigrate Design Center, Georgia Institute of Technology, Atlanta, Georgia, United States

Adaptive technology for older adults with cognitive impairment is typically designed around individual preference, yet most of this population lives and cooks with a spouse or family member. This paper examines a co-design workshop in which four dyads and two individuals (N=10) built kitchen cabinet designs from twenty-one options across five features. Thematic analysis of thirty selections identifies recurring patterns: visual access retained through enclosure rather than open shelving, physical effort treated as a household concern, and automation accepted when predictable. Structured analysis of ten interaction episodes shows how some patterns were negotiated in practice, including care partners contributing embodied constraints distinct from the primary participant's, and disagreements resolving through documented deliberation. Together, the two analyses show that adaptive technology preferences are not always individual properties but can be shaped through household interaction. We offer candidate design implications for facilitation protocols and collaborative systems supporting shared decision-making in aging-in-place contexts.

## 1 INTRODUCTION

Older adults with cognitive impairment increasingly rely on adaptive technology to remain in their own homes, and the kitchen is one of the most demanding rooms in which to do so: it requires locating objects out of sight, sequencing multi-step tasks, and retrieving items whose position may change [1, 2]. Design guidance for this population often centers on the preferences of the individual with the diagnosis. Many people with cognitive impairment, however, live with a spouse or family member who shares the kitchen and participates in daily household decisions [17], and technology adoption in these households is not necessarily made by one person alone.

The participatory design literature has documented that care partners are usually present at co-design sessions, but it has rarely examined what happens when they and the person with cognitive impairment negotiate a shared decision. Existing co-design work with this population tends to treat the presence of a second person as a logistical or supportive feature of the session rather than as a social process worth analyzing on its own terms. When care partners are discussed, they are more often framed as proxies who speak on behalf of the person with cognitive impairment, or as validators who confirm a choice, than as co-designers with their own embodied stakes in the outcome.

This framing leaves an empirical question open. A workshop that asks participants to select from a defined set of design options can tell us what older adults with cognitive impairment and their care partners choose, and it can tell us the rationales they give. It cannot, by itself, tell us whether those choices were reached individually or through negotiation, nor whether the negotiation itself shaped what was chosen. Answering that question requires looking at the same data in two ways: once across the full set of choices to establish what was chosen and why, and once inside specific episodes of interaction to see how a choice was actually reached.

This paper reports both. We conducted a participatory co-design workshop with four dyads, each comprising an older adult with cognitive impairment or cognitive concerns and a care partner, and two individual participants with self-reported cognitive concerns (N=10) at a community-based cognitive wellness program in the southeastern United States. Participants used a structured card-based method to build an adaptive kitchen cabinet across five functional features, each offering three to five design options spanning low to high technology, recorded on a completed board for each group (Figure 2). We analyze the resulting thirty selections in two ways. A reflexive thematic analysis of the complete dataset identifies the design patterns and boundary conditions that recur across all six groups. A structured analysis of ten purposively selected interaction episodes then shows how some of those patterns were negotiated between the people who produced them.

This dual-analysis design lets us ask three questions that neither analysis could answer alone. RQ1: What design configurations and underlying rationales emerge when older adults with cognitive impairment or cognitive concerns and their care partners co-design adaptive kitchen storage? RQ2: How do dyadic interactions, including agreement, disagreement, participant agency, and care-partner embodiment, shape those design outcomes? RQ3: What do these preference and interaction patterns imply for the design of adaptive technologies in shared domestic environments?

We make three contributions, corresponding to RQ1 through RQ3. First, we show that visual access, physical effort, and automation acceptance in kitchen storage design follow recurring patterns across a full purposive sample, several of which qualify or extend prior experimental findings about this population. Second, we show that a subset of these patterns can be traced to specific interactional mechanisms, including care partners contributing constraints and embodied needs distinct from the primary participant's, and disagreements that were worked through rather than avoided. Third, we argue that these findings together support the possibility that adaptive technology preference can be constructed through household interaction rather than being only a fixed attribute of one person, with direct implications for co-design facilitation and for systems that support shared decision-making in the home.

## 2 RELATED WORK

### 2.1 Technology Preference and Adaptive Design in Cognitive Aging

Mild cognitive impairment affects a substantial share of older adults and is associated with difficulty performing instrumental activities of daily living, including meal preparation [1]. Design responses to these difficulties generally fall into three categories: visual strategies that remove or glaze enclosures to address out-of-sight forgetting, memory strategies that externalize prospective memory through labels or recognition cues, and automation strategies that offload initiation of an action to a sensor. Each has experimental support in isolation [2, 3, 4], but how these strategies are received when embedded in an actual kitchen, chosen from among competing alternatives, and evaluated by the people who would use them, is comparatively underexamined. Prior experimental work has shown that open shelving reduces search time, cognitive load, and physical effort for older adults with cognitive impairment [5], yet whether people actually want open shelving in their own homes is a separate question from whether it improves performance in a controlled setting.

Kitchen-specific design guidance for this population has largely been derived from task analysis and expert principles rather than from participatory input. Prior work has catalogued the specific kitchen tasks that become difficult with dementia, including sequencing, object recognition, and recovery from interruption [22], and has proposed design principles for a "cognitive kitchen" built around similar task breakdowns [23]. Participatory methods have been used successfully with this population in adjacent domains, including photo-based co-design of the built environment [13] and digital storytelling interventions that used iterative co-design and usability testing with people with mild cognitive impairment [20]. These studies establish that people with cognitive impairment can meaningfully participate in design activities; what they do not establish is how a shared decision, rather than an individual response, gets made when a second household member is also part of the session.

### 2.2 Shared Domestic Environments and Care-Partner Participation

Kitchens are shared spaces, and cabinet requirements are set not only by the person with a diagnosis but by whoever else uses the room. Family caregivers have been characterized as boundary actors who exercise decision-making authority extending well beyond their formally recognized role, mediating between the person they support, other family members, and outside institutions [6]. Applied to a co-design context, this suggests that care partners may occupy an intermediating position in the design process, shaping what gets voiced, validated, and selected, rather than the purely supportive role they are usually assigned in the literature. Technology acceptance frameworks for aging in place already extend beyond usability to include social influence and the meaning a device carries in the home [7], often described under the heading of stigma [8], though that term is rarely broken into its component concerns. A further tension follows from designing for an individual user within a household: a feature that suits one occupant's needs may inconvenience another's [9], and existing co-design studies with this population have not systematically examined how that tension is resolved when it arises.

This tension echoes a broader concern in CSCW research on health-related technology in the home, where the unit that actually adopts and sustains a technology is often a dyad or household rather than an individual patient. Telehealth companionship technologies for aging in place have been shown to require ongoing relational work between the person receiving care and the person providing it, work that is not captured by treating either party as a passive user [18]. Conversational-agent systems designed explicitly for dyads of older adults with mild cognitive impairment and their care partners have likewise found that the two parties bring distinct needs and interaction patterns to a shared system, needs that a single-user design would not surface [19]. CSCW as a field has historically been concerned with how technology mediates and is shaped by cooperative work among multiple people [21]; a kitchen co-design session in which a primary participant and a care partner jointly select an adaptive feature is a small-scale instance of exactly this problem, though it has not been analyzed in these terms.

### 2.3 Co-Design as Negotiation and Decision-Making

Design is fundamentally a decision-making process in which power and participation are distributed unevenly across participants and moments, and understanding co-design requires attending to who makes which decisions, under what conditions, and with what consequences for others [10]. This analytical orientation has rarely been applied empirically to co-design sessions involving a person with cognitive impairment and a care partner, where the presence of a second decision-maker introduces a relational layer that individual-focused methods cannot capture. Reviews of co-design projects for aging in place have found that few studies reach the level of specific artifact configuration [12], leaving open how a concrete design decision, rather than a general design direction, gets made when more than one household member is involved. Iterative design work with care partners in adjacent domains, including context-aware meal preparation assistance, has shown that care partners can surface considerations that the person with cognitive impairment could not articulate alone, and that collaborative design sessions produce specifications that better reflect household realities than individual participation does [11]. This orientation toward design as a decision-making process builds on a longer tradition of participatory design theory concerned with co-creation as a distinct landscape of design activity [24], and with the history of inclusive design as a practice that treats accommodation for a range of users as a mainstream design concern rather than a specialism [25]. What remains unexamined is the mechanism connecting the presence of a care partner to a specific design outcome: whether a selection reflects one party's preference with the other's assent, an averaged compromise, or something else entirely that neither party would have proposed alone.

Taken together, this literature leaves three related gaps underexamined. First, how cognitive-impairment and care-partner dyads negotiate design preferences in real time, including how disagreements are resolved, has received limited empirical attention. Second, the role of care partners as agents with their own physical constraints and household concerns, distinct from their caregiving role, has not been systematically examined in design contexts. Third, stigma management in this population's assistive technology research is typically treated as an individual attribute, despite evidence that it can operate through social and relational mechanisms shaped by who is present during a design decision. This paper addresses these gaps through a co-design workshop analyzed at two levels of grain: across the complete set of design choices, and within specific episodes of interaction that produced some of them.

## 3 METHODS

### 3.1 Design and Setting

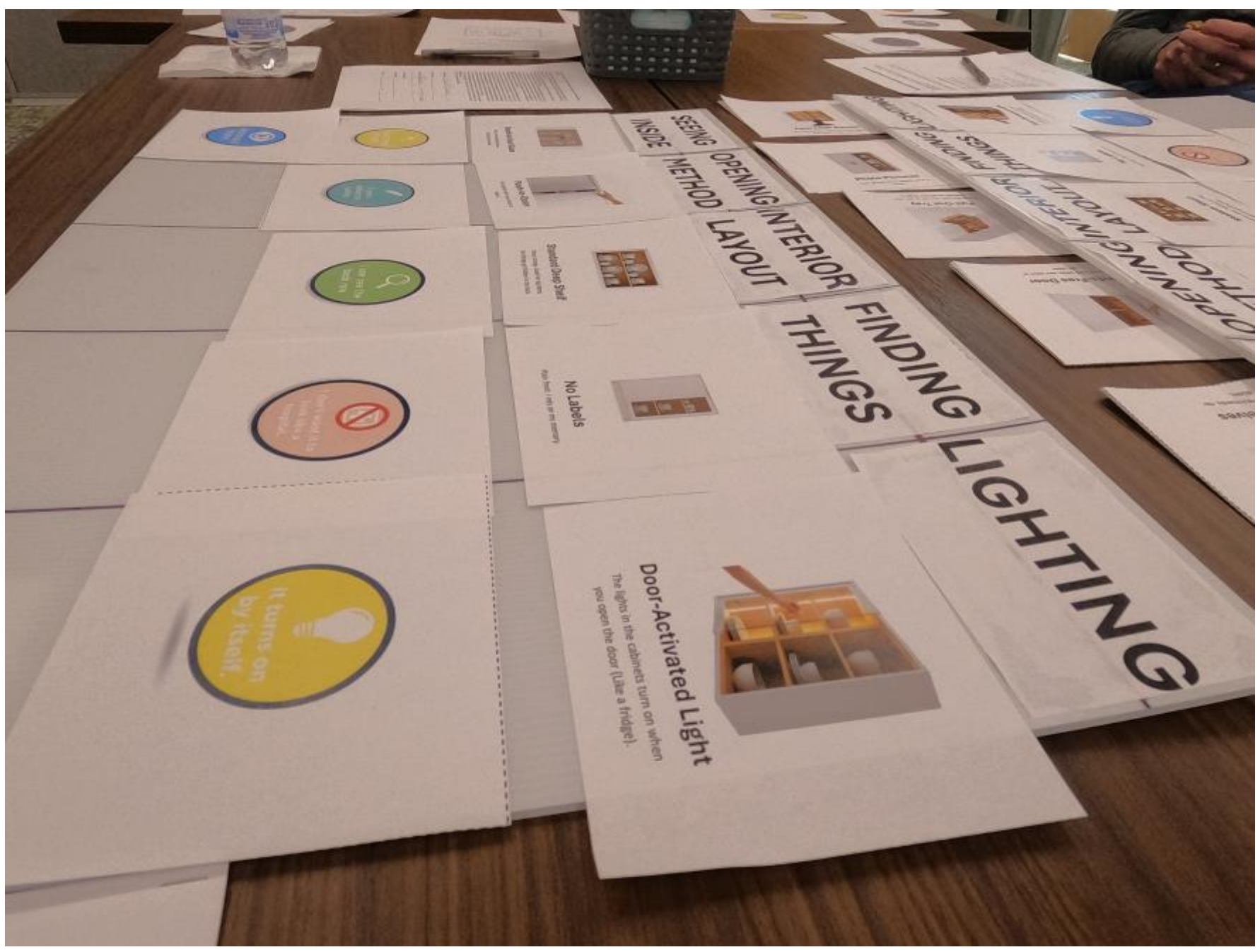

Figure 1. Workshop materials in use during a session.

The workshop used a structured card-based method to elicit and examine preferences for adaptive kitchen cabinet features. Cabinet design was decomposed into five functional features: seeing inside, opening method, interior layout, finding things, and lighting. Each feature was represented by a dedicated station offering three to five option cards spanning low, mid, and high technology, for twenty-one options in total. Participants attached up to two pre-printed reason cards to each selection to record their rationale, with open-ended cards available when no pre-printed option matched a participant's intent.

The room was arranged into three pod sections formed by joined tables, each accommodating one or two groups, with a central buffet area holding all option and reason cards (Figure 1). Option cards paired plain-language descriptions with visual concepts, following prior work on photo-based co-design methods for this population [13]. Three facilitators managed the session: one served as lead presenter and overall facilitator, and two provided table-level facilitation and documentation. The workshop ran as a single two-hour session, of which forty-five to sixty minutes constituted the core card-selection activity. Observers recorded feature selections, reason card placements, and free-text field notes at each station, along with a quick-coding scheme flagging hesitation, enthusiasm, inter-participant disagreement, technology interest, and fatigue. In practice, only one observer recorded these codes at a given station, so counts from the code scheme are not reported; the qualitative analyses below rest on the free-text notes. Sessions concluded with a brief satisfaction rating and compensation. Following preliminary analysis, a forty-one-minute recorded appraisal session was held with two clinical specialists affiliated with the recruiting program, an occupational therapist and a physical therapist, structured around a walkthrough of the completed boards and the research team's preliminary interpretations. This session is reported as informed commentary rather than independent validation, since the clinicians responded to interpretations the team had already formed. Study procedures were approved by the first author's institutional review board, and written informed consent was obtained from all participants.

### 3.2 Participants

Participants were purposively recruited through a community-based cognitive wellness program and an associated public event in the southeastern United States. Inclusion criteria for primary participants were age fifty or older, with or without a clinical diagnosis of cognitive impairment, and regular kitchen use. Care partners were spouses assisting with daily activities. The final sample comprised ten individuals: six primary participants and four care partners, forming four dyads and two individuals participating without a partner. Among primary participants, three reported a diagnosis of mild cognitive impairment, one a diagnosis of Alzheimer's disease, and two no formal diagnosis but self-reported cognitive concerns. Diagnoses are self-reported throughout. Full characteristics appear in Table 1.

Table 1. Participant demographic characteristics (N=10). Diagnoses and memory change are self-reported.

| Characteristic | Primary Participants (n=6) | Care Partners (n=4) | Total (N=10) |
|---|---|---|---|
| Age, M (SD) | 72.3 (4.1) | 71.0 (2.4) | 71.8 (3.4) |
| Gender (M/F) | 4/2 | 1/3 | 5/5 |
| Race/ethnicity | White (3), Black (1), Other (2) | White (4) | White (7), Black (1), Other (2) |
| Relationship to primary participant | -- | Spouse (4) | -- |
| Same household | -- | 4 (100%) | -- |
| Assisting duration | -- | <1 year (2), 1-3 years (2) | -- |
| Kitchen use, daily | 6 (100%) | 4 (100%) | 10 (100%) |
| Primary kitchen challenge | Clutter (4), Remembering locations (1), Other (1) | High shelves (3), Low cabinets (1) | Clutter (4), High shelves (3), Remembering (1), Low cabinets (1), Other (1) |
| Diagnosis | MCI (3), Alzheimer's disease (1), None (2) | None (4) | MCI (3), Alzheimer's disease (1), None (6) |
| Self-reported memory change | Yes (2), No (2), Not recorded (2) | N/A | N/A |

### 3.3 Positionality

The first author has a background in architecture and building construction. He designed the workshop instruments and card sets, facilitated three of the six sessions, was present for and later coded the deliberations of groups he facilitated, and conducted the appraisal session. The first author's earlier experimental work [5] had reported benefits of open shelving for this population, which created scope for confirmation bias in interpreting a workshop where open shelving was rarely chosen. To address this, the reflexive thematic analysis reported below is supported by a complete codebook and coding matrix covering every selection, supplied as supplementary material so that readers can weigh the interpretation against the underlying evidence. For the structured episode analysis, the first author was one of three coders. The separate episode analysis included coders with differing levels of involvement in the study: an independent coder with a design background and no prior involvement, and a third coder with a design background who had participated in data collection.

### 3.4 Analysis of Design Selections and Rationales

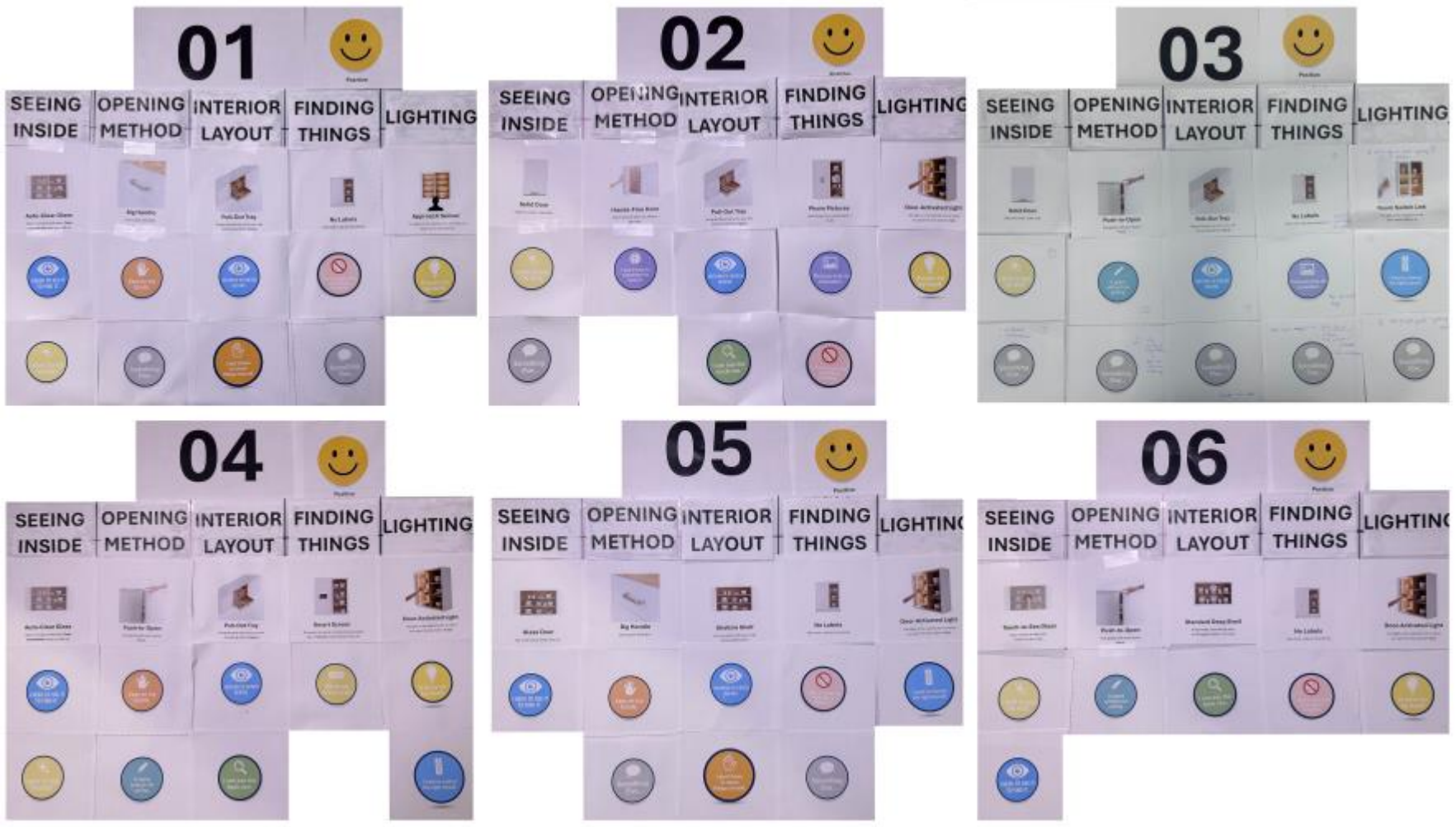

Figure 2. Completed boards from each of the six groups.

The complete dataset of thirty selections, six groups across five stations, was analyzed using reflexive thematic analysis [14, 15], which treats coding as an interpretive act by a situated analyst rather than a measurement to be verified against a second rater. No inter-coder reliability statistic is reported for this analysis, and none would be appropriate to it. Transparency is instead provided through an evolving codebook recording code definitions and boundary decisions, together with a coding matrix covering every selection and its supporting evidence from the reason cards and observer notes, both supplied as supplementary material. Coding proceeded through familiarization with the boards and field notes, generation of initial codes grounded in stated rationales, refinement of code boundaries, clustering into candidate themes, and review against the full dataset. The final codebook contains twenty-three codes describing categories of rationale such as visual access, concealment, biomechanical effort, memory scaffolding, and household considerations. Codes were developed by the first author and discussed iteratively with a co-author acting as a critical friend, a role intended to interrogate the coherence of emerging themes rather than to verify them independently.

### 3.5 Interaction-Episode Analysis

A second, methodologically distinct analysis examined how design decisions were reached in specific moments of interaction. From the thirty selections, ten episodes were purposively selected for their analytic richness: each showed a documented disagreement or hesitation code, substantive observer notes describing interaction dynamics, or reason-card combinations suggesting competing rationales within the group. These ten episodes were not selected to be representative of the full dataset in frequency terms, and the findings from this analysis describe the mechanisms visible within these episodes rather than estimating how often such mechanisms occur across all thirty selections. This selection criterion intentionally privileges episodes with richer interactional documentation and therefore likely overrepresents disagreement, hesitation, and complex reasoning relative to the full set of thirty selections.

Each episode was independently coded by three raters along four dimensions directly observable from the record: deliberation depth, presence of disagreement, how the selection was resolved, and the type of reasoning invoked. Two further dimensions, who initiated the selection and who drove the final decision, were coded only where observer notes provided unambiguous attribution and are reported narratively rather than in the summary table, since attribution was not consistently documented across episodes. The three coders were the first author, who served as workshop facilitator and brought contextual knowledge of the sessions; an independent coder with a design background and no prior involvement in the study; and a third coder with a design background who had participated in data collection, providing familiarity with the workshop setting without involvement in the study's analytical framing. Coding followed independent scoring by all three raters, following established qualitative practice of independent coding paired with structured consensus discussion [16], followed by discussion of disagreements to reach consensus. The unit of agreement was the episode-dimension: each of the ten episodes scored on each of the four dimensions, for forty episode-dimension classifications in total. Thirty-six of these forty classifications showed unanimous agreement across all three coders before discussion, for an initial agreement rate of ninety percent; the remaining four were resolved through consensus discussion, reaching full agreement on all forty. This agreement procedure assesses consistency across coders for these four structured classifications. It does not validate the reflexive themes reported in Section 3.4, which follow a different analytic logic and were not designed to be independently reproducible in the same sense.

## 4 RESULTS

Table 2 summarizes selections across all thirty placements; Table 3 summarizes the ten interaction episodes examined at the mechanism level. Figure 3 groups the rationale codes underlying these patterns into the six themes identified in the complete dataset.

Table 2. Feature selections by station (n=30 selections). Each group made one selection per station; five of twenty-one options offered were selected by no group.

| Station | Option | Technology Level | Groups Selecting (of 6) |
|---|---|---|---|
| Seeing Inside | Auto-Clear Glass | High (passive) | 2 |
| Seeing Inside | Solid Door | Low | 2 |
| Seeing Inside | Touch-to-See Glass | High (active) | 1 |
| Seeing Inside | Glass Door | Mid | 1 |
| Seeing Inside | Open Shelves | Low | 0 |
| Opening Method | Push-to-Open | Mid | 3 |
| Opening Method | Big Handle | Low | 2 |
| Opening Method | Hands-Free Door | High | 1 |
| Opening Method | Always Open | Low/Mid | 0 |
| Interior Layout | Pull-Out Tray | Mid | 4 |
| Interior Layout | Standard Deep Shelf | Low | 1 |
| Interior Layout | Shallow Shelf | Low | 1 |
| Finding Things | No Labels | Low | 4 |
| Finding Things | Photo Pictures | Low | 1 |
| Finding Things | Smart Screen | High | 1 |
| Finding Things | Text Labels | Low | 0 |
| Finding Things | Color Zones | Low | 0 |
| Lighting | Door-Activated Light | Mid | 4 |
| Lighting | Approach Sensor | High | 1 |
| Lighting | Room Switch Link | Low | 1 |
| Lighting | High-Contrast Edges | Low | 0 |

Table 3. Ten purposively selected interaction episodes coded for four structured, observable dimensions. Frequencies describe these episodes only.

| Episode | Deliberation Depth | Disagreement | Resolution | Reasoning Type | Changed |
|---|---|---|---|---|---|
| Dyad1_S1 (Auto-Clear Glass) | High: two competing reason cards | Yes, implicit | Technology-mediated compromise | Mixed (functional + aesthetic) | Yes |
| Dyad1_S4 (No Labels) | Moderate: glaucoma comorbidity noted | No conflict | Unchallenged selection | Mixed (stigma + functional comorbidity) | No |
| Dyad2_S2 (Hands-Free Door) | High: care-partner preference noted | Yes, explicit | Primary participant selection prevailed | Functional (prospective memory) | Yes |
| Dyad2_S3 (Pull-Out Tray) | High: care-partner physical need noted | No conflict | Unchallenged selection | Safety (care-partner embodied) | Yes |
| Dyad3_S1 (Solid Door) | High: lengthiest deliberation of Seeing Inside episodes | Yes, explicit | Primary participant framing prevailed | Mixed (functional + aesthetic) | Yes |
| Dyad3_S4 (No Labels) | High: explicit disagreement, provisional framing | Yes, explicit | Primary participant selection prevailed | Social-stigma (progressive) | Yes |
| Dyad6_S1 (Glass Door) | High: debated | Yes, explicit | Compromise selection | Mixed (visibility + protection) | Yes |
| Dyad6_S4 (No Labels) | Low: easy decision noted | No conflict | Unchallenged logical inference | Functional (cross-category) | No |
| Single4_S2 (Push-to-Open) | Moderate: reason change noted | N/A (solo) | Solo-resolved | Functional (drifted reasoning) | Yes |
| Single5_S4 (No Labels) | Low: straightforward noted | N/A (solo) | Solo-resolved | Social-stigma (visitor-oriented) | No |

### 4.1 Visual access was retained, but exposure was negotiable

Across the complete dataset, open shelving, offered as a low-technology option at the seeing-inside station, was selected by no group. Four of the six groups instead selected a glass variant: two chose an auto-clearing glass that changed opacity, one a touch-activated glass, and one a conventional glass door. The remaining two groups chose a solid door. Reason cards for the glass selections combined visual access with concealment: three of the four groups that chose a glass variant placed reason cards for both seeing and hiding at the same station, and one description characterized the appeal as a kitchen that feels open while the glass itself stays opaque enough to conceal the contents. The pattern was not uniform. In one group, the primary participant argued against visual accessibility on sensory grounds, describing himself as claustrophobic and reporting that his brain no longer separated visual information the way it once had, so that seeing many items at once would be harmful rather than helpful.

The interaction-episode analysis shows how some of these outcomes were reached. In one dyad, the couple selected a glass door specifically because it retained a closing barrier that protected the contents from splashes, after the care partner had expressed a preference for closed cabinets and the primary participant a preference for visibility. The observer notes record the care partner's reasoning directly: closed cabinets felt more protected, while the primary participant wanted something easy to see. The group settled on a glass door because it offered a door that closed while still allowing visibility, a resolution that integrated concerns neither party's starting position had fully addressed. A second, differently resolved episode illustrates the exception rather than the trend: this group's primary participant, an architect by training, argued against visibility on sensory grounds, describing himself as claustrophobic and reporting that his condition made seeing many items at once harmful rather than helpful. Observer notes record this as the lengthiest deliberation of any seeing-inside selection, and the group's final choice of a solid door followed the primary participant's framing rather than a compromise. This episode shows that the boundary condition identified in the full dataset, that visual accessibility is not universally preferred, was itself the product of one participant's reasoning being taken seriously and prevailing within his dyad's deliberation.

### 4.2 Physical accessibility was a household requirement, not an individual one

Four of six groups selected a pull-out tray at the interior layout station, the most frequently selected option offered at that station. Reason cards indicating that nothing is hidden behind the shelf were placed at five of the six selections, and cards indicating visibility of the back row at three. In two groups, the physical constraint motivating the selection belonged to the care partner rather than the primary participant.

One of these two episodes was selected for the interaction-level analysis. The care partner explained that her own joint problems made lower cabinets difficult to access, describing getting down on one knee as the hardest part of retrieving items, and reported having long admired pull-out trays for exactly this reason. The primary participant's stated rationale for the same selection concerned visibility rather than reach. Both rationales were recorded on the board, and the observer notes attribute the physical-effort argument specifically to the care partner. This episode illustrates a pattern the complete dataset cannot show on its own: an adaptive feature was selected to address a physical constraint that belonged to the person without the diagnosis, motivated by concerns about her own body rather than the cognitive needs of the person she was supporting. Design guidance that treats the primary participant as the sole intended user of an adaptive feature would miss this dimension of the household's requirement entirely.

### 4.3 Memory support was negotiated against identity and progression

No labels was the most frequently selected option at the finding-things station, chosen by four of six groups. A reason card expressing reluctance for the cabinet to resemble a hospital was placed at four of the six selections. Two groups gave rationales the pre-printed card set did not anticipate: one cited a co-occurring visual impairment that made any text difficult to read regardless of labeling, and another treated labeling as redundant given an already-transparent cabinet front.

The same visible outcome, no labels, arose from different reasoning in different groups, and the interaction-episode analysis helps distinguish those reasons rather than collapsing them into a single explanation. In one dyad, the selection was reached after documented disagreement, and observer notes record that the primary participant's preference prevailed. The group's stated rationale was explicitly provisional: the participant had been weighing plain fronts against photographic cues, and chose the plain front for present use while stating that photographs could be added later if his condition advanced, reasoning that it is easier to add a memory cue afterward than to remove one. In a different, individual case with no care partner present, the participant's stated reason for avoiding labels was direct and unmediated by any negotiation: not wanting visitors to sense that someone in the household had cognitive difficulties. Both cases produced the same visible board outcome, but one reflects a negotiated position about anticipated future decline and the other a direct statement about present social exposure. Treating either as evidence of stigma alone would erase the distinction between them.

### 4.4 Automation was acceptable when predictable

Door-activated lighting was selected by four of six groups, more than any other lighting option, and reason cards indicating automatic activation were placed at four of those selections. Acceptance did not track the overall degree of automation, however. Two groups selected the passive, automatic light in preference to an approach sensor that would activate from a distance, one citing a negative prior experience with an automatic trash bin that opened whenever he walked past regardless of need, and the other stating plainly that they did not want a light triggered by anyone simply approaching. A third group selected a manual switch instead, arguing that the room, not the cabinet, was the appropriate scale for a lighting intervention. Meanwhile, two groups elsewhere in the dataset selected auto-clearing glass, a feature that similarly changes state without the user initiating an action.

Read together, these choices show that presence-triggered operation was accepted in some designs and declined in others, and that what varied was not the presence of automation but its predictability: whether activation could be anticipated, whether it occurred when unwanted, and whether it required learning something new. High-technology options accounted for only one-fifth of all selections across the dataset, so interest in automation in this population should not be read as a general appetite for technology. In this dataset, acceptance tracked specific properties of how a given automated feature behaved rather than the presence of automation as a category.

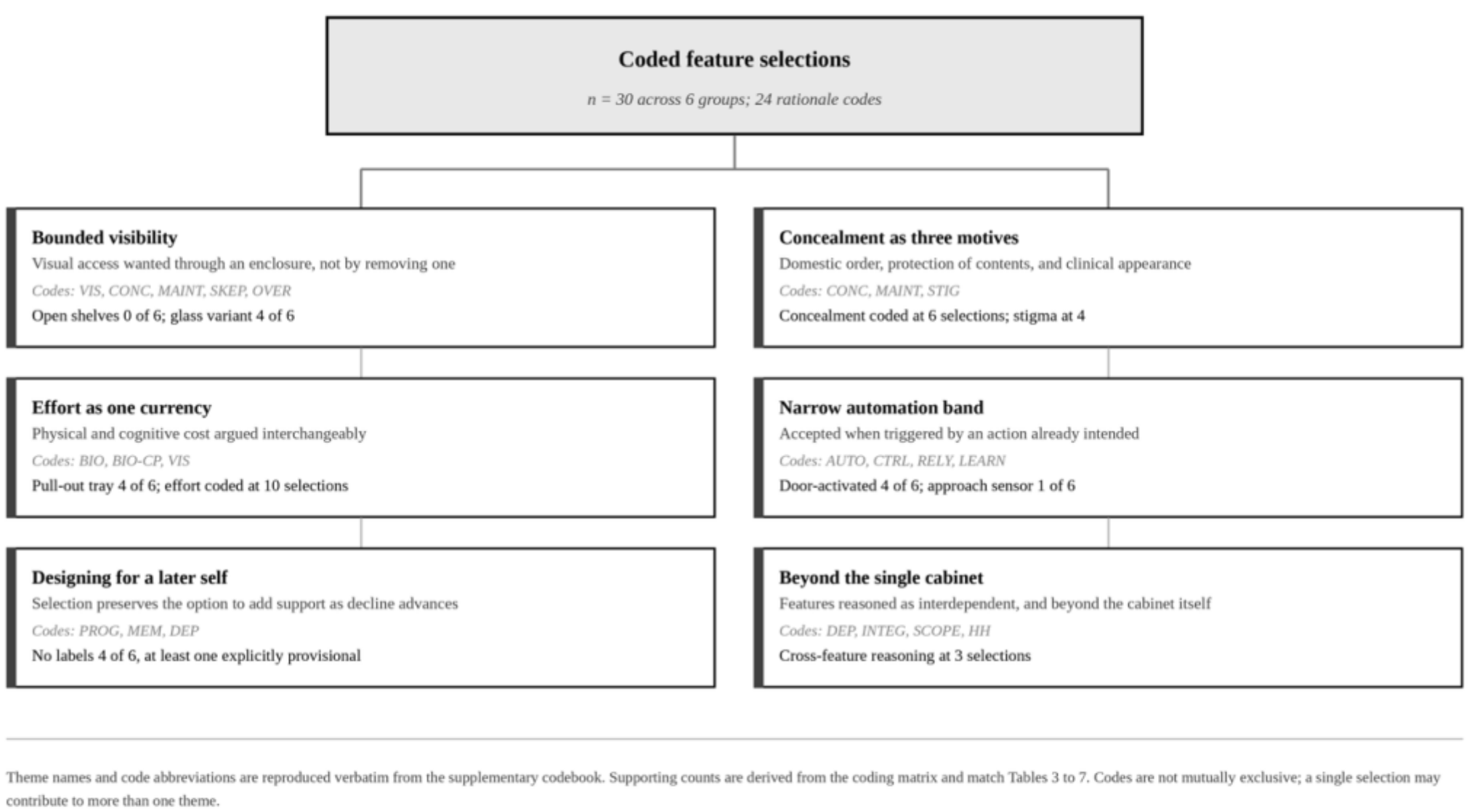

Figure 3. Rationale codes grouped into the six themes from the complete dataset.

### 4.5 Preferences changed across people, configurations, and anticipated futures

Three further patterns cut across the stations already discussed. First, whose needs mattered varied by selection: in two of the episodes described above, the constraint driving a choice belonged to the care partner rather than the primary participant, and in a further case a group's opening-method choice was shaped by accommodating visiting grandchildren rather than either adult in the dyad. Second, one feature's value sometimes depended on another feature already chosen: a group that had selected a glass door reasoned that labels were redundant because contents would already be visible, while noting that an opaque enclosure would have made labeling necessary again. Third, at least one selection anticipated a future rather than a present need, as described in Section 4.3, where a participant chose a plain front now on the understanding that photographic cues could be added later, but not the reverse.

None of these three patterns is visible from a single station's frequency count alone. They emerge only when a selection is read alongside the reasoning that produced it and, in the two episodes analyzed at the interaction level, alongside the deliberation that shaped it. Table 4 draws together five candidate design implications from the complete dataset, each paired with a boundary condition or exception that qualifies it.

Table 4. Design implications and candidate boundary conditions. Offered as candidates from a small purposive sample, not validated design requirements.

| Design implication | Evidence in this study | Boundary condition or exception |
|---|---|---|
| Visual access may be more acceptable when delivered through an enclosure | 4 of 6 selected a glass variant; open shelves selected by no group | 2 selected solid doors; one primary participant described visual access as overstimulating |
| Pull-out storage addresses visibility and reach together | 4 of 6 selected pull-out trays; visibility and reach cited together | One group selected a deep shelf for storage capacity |
| Memory supports should remain discreet and additive | 4 of 6 selected no labels; one selection explicitly provisional | Photo cues and a smart screen were also selected |
| Automation requires predictable activation and low learning demand | Door-activated lighting selected by 4 of 6; unwanted activation raised in two groups | One group selected the approach sensor; two selected auto-clear glass |
| Storage must serve the household, not one diagnostic user | Care partner constraints shaped 2 selections; grandchildren a third | Household needs can conflict; no general or formal design mechanism for reconciling such conflicts was tested here |

## 5 DISCUSSION

This study set out to examine what adaptive kitchen technologies older adults with cognitive impairment and their care partners choose, and how those choices come about. The complete selection dataset identifies recurring design conditions across six groups: enclosed but transparent storage was preferred to fully open shelving, physical effort was treated as a shared household concern rather than an individual one, and automation was accepted when its behavior was predictable rather than avoided outright. The interaction episodes illuminate how some of these conditions were negotiated in practice, showing documented moments in which a care partner's own physical needs, a primary participant's professional reasoning, or an anticipated future decline shaped a selection that neither the frequency data nor a single participant's account would fully explain. We draw three contributions from this combination.

First, adaptive technology preference in this context is better understood as relationally constructed than as a fixed attribute waiting to be elicited from an individual. The glass-door compromise in Section 4.1 was not a data point that could have been predicted from either party's stated position alone; it existed only because both positions were voiced and reconciled within the session. This connects to a broader concern in CSCW research on health-related technology in the home, where relation work between the person receiving care and the person providing it has been shown to shape technology use in ways that treating either party as an independent user cannot capture [18]. The dyadic negotiation documented here extends that concern from ongoing technology use to the design decision itself: the preference a system would need to elicit does not exist in a stable form prior to the interaction that produces it. A system, survey, or facilitation protocol that samples one household member at a time risks missing the outcome that actually gets built.

Second, care partners in this dataset acted as embodied co-users rather than proxies. The pull-out tray selection in Section 4.2 was driven by the care partner's own joint pain, not by an assessment of the primary participant's cognitive needs. Existing HCI and CSCW accounts of care partners have largely cast them as boundary actors who mediate between the person they support and outside systems and institutions [6], a framing that positions their contribution as translation or advocacy on someone else's behalf. The evidence here suggests a role that framing does not fully cover: a care partner whose own body has stakes in the design outcome, independent of any mediating function. This distinction matters for design practice: an intervention framed as assistive technology for the person with a diagnosis may in fact be serving, or failing to serve, the person who does not have one. Household-level design requirements cannot be inferred from the primary participant's profile alone.

Third, adaptation in this population is both configurational and temporal. The configurational evidence is clearest in Section 4.5: a group that had selected a glass door treated labels as redundant because contents were already visible, while noting that an opaque enclosure would have made labeling necessary again, so the value of one feature changed depending on which other feature had already been chosen. The temporal evidence appears in Section 4.3, where at least one preference was explicitly built to change as a condition progresses. Static design specifications, calibrated once to a stated preference, may not remain fit for purpose as either the household's configuration of features or the primary participant's condition shift. Automation predictability, discussed in Section 4.4, is a related but separate design implication rather than direct evidence for this third contribution.

Participant agency, disagreement, and identity management functioned as mechanisms supporting these three contributions rather than as separate findings in their own right. Across the coded episodes, primary participants exercised substantial and at times decisive agency in disagreements: where a documented conflict occurred, the primary participant's selection or framing prevailed more often than a compromise or the care partner's preference did. Disagreement itself was not a failure of the design process; richer deliberation episodes sometimes produced outcomes that incorporated concerns voiced by both members, though the coding scheme does not support a general claim that more deliberation produced better-supported outcomes. And concealment, discussed in Section 4.3, carried at least three distinct meanings in this dataset: domestic tidiness, protection of contents, and discretion about cognitive status. Only the last of these constrains assistive-technology design in a way that tidiness or protection do not, and conflating them risks treating every closed-front cabinet as evidence of stigma when the same visible choice may reflect an entirely different concern.

These findings carry implications for the design of both facilitation protocols and collaborative computing systems intended to support household decision-making. Facilitators working with dyads where one member has cognitive impairment should avoid prematurely suppressing disagreement and may benefit from supporting its constructive resolution, since several of the richest episodes in this dataset resolved through disagreement rather than around it.

For system builders, these findings point toward specific design directions rather than a single unified recommendation. Systems that elicit preferences from a single household member, or that aggregate individual responses into one household profile, risk missing the process through which a shared decision is actually reached. The episodes analyzed here suggest concrete alternatives: supporting simultaneous input from both household members rather than sequential individual elicitation; preserving a disagreement as a recorded state rather than resolving it into an average; recording whose rationale produced a given selection, since the same board outcome in this dataset sometimes concealed distinct and even conflicting reasons; treating a stated preference as negotiable and revisable rather than fixed at first elicitation, consistent with the provisional labeling choice in Section 4.3; and allowing a household's configuration of features to be adjusted over time as circumstances change, rather than fixed once at setup. Retaining the provenance of a preference, not only its current value, may help a system distinguish a jointly agreed selection from one reached through compromise or unresolved tension.

## 6 LIMITATIONS

The sample comprised ten participants in six groups recruited from a single community-based program, and the descriptive patterns reported here characterize this purposive sample rather than a representative population. Recruitment through a cognitive wellness program favors households already engaged with support services. Diagnostic status was self-reported and heterogeneous: three primary participants reported mild cognitive impairment, one Alzheimer's disease, and two no formal diagnosis alongside self-reported cognitive concerns; findings should not be attributed to a diagnostically uniform group, and the ten interaction episodes analyzed at the mechanism level should be read as demonstrating that these negotiations occur and how, not as estimating how often they occur across the full range of households this population represents.

The option set itself shaped some results: the finding-things station offered four low-technology options and only one high-technology option with no mid-technology alternative, so the low-technology outcome at that station is partly a structural feature of the instrument rather than solely a finding about preference. The reflexive thematic analysis was conducted primarily by a single analyst who also designed the instruments and facilitated three of the six sessions; the codebook and coding matrix are supplied as supplementary material so that this interpretation can be checked against the underlying evidence. The structured episode analysis addresses part of this concern through independent multi-coder agreement, but that agreement applies only to the four observable classifications it covers and does not extend to the reflexive themes drawn from the complete dataset. All groups rated their designs positively in a single session with researcher presence and compensation, so these ratings should not be read as evidence of eventual adoption. The clinical appraisal session followed a researcher-led walkthrough in which preliminary interpretations were presented before questions were asked, and both clinicians were affiliated with the recruiting program; their contributions are reported as informed commentary rather than independent corroboration. Finally, the interaction-episode analysis relies on observer field notes, quick codes, reason cards, and completed boards rather than verbatim transcripts of workshop dialogue. This supports claims about documented negotiation and its recorded outcome, but does not support fine-grained claims about conversational sequence, turn-taking, or moment-to-moment power dynamics, which would require conversation-analytic data this study did not collect.

## 7 CONCLUSION

When older adults with cognitive impairment and their care partners were given a structured set of adaptive kitchen options, no group chose open shelving, while four of six selected some form of glass enclosure and two chose a solid door. Where we could examine the process behind a choice, some apparently simple final selections concealed distinct household rationales and negotiation processes that a single frequency count would not reveal, though not every changed selection reflected a negotiated compromise rather than one party's preference prevailing. Preference in this context is not simply a property of the person with a diagnosis. In shared domestic contexts, it can be something a household builds together, and design methods, facilitation practices, and support systems that treat it as fixed and individual will miss part of what they are meant to serve.